\documentclass[twocolumn,preprintnumbers,amsmath,amssymb,prl]{revtex4-1}

\usepackage{amsmath} % Advanced math typesetting
\usepackage[utf8]{inputenc} % Unicode support (Umlauts etc.)
\usepackage[english]{babel} % Change hyphenation rules
\usepackage{graphicx} % Add pictures to your document
\usepackage{listings} % Source code formatting and highlighting
\usepackage{soul}
\usepackage{color}

\begin{document}

\title{Effective potentials induced by self-assembly of patchy particles}
\author{Nicol\'as Ariel Garc\'ia$\S$} 
\author{Nicoletta Gnan$\S$}\email[]{nicoletta.gnan@roma1.infn.it}
\author{Emanuela Zaccarelli} \email[]{emanuela.zaccarelli@cnr.it}
\affiliation{CNR-ISC UOS {\em Sapienza} and Department of Physics,``Sapienza" University of Rome, Piazzale A. Moro 2, 00185 Roma, Italy }
\affiliation{$\S$ These authors contributed equally to this work.}
\date{\today}

\begin{abstract}
Effective colloid-colloid interactions can be tailored through the addition of a complex cosolute. Here we investigate the case of a cosolute made by self-assembling patchy particles. Depending on the valence, these particles can form either polymer chains or branched structures. We numerically calculate the effective potential $V_{eff}$ between two colloids immersed in a suspension of reversible patchy particles, exploring a wide region of the cosolute phase diagram and the role of valence. In addition to well-known excluded volume and depletion effects, we find that, under appropriate conditions, $V_{eff}$ is completely attractive but shows an oscillatory character.
In the case of polymerizing cosolute, this results from the fact that chains are efficiently confined by the colloids through the onset of  local order. This argument is then generalized to the case of particles with higher valence, under the condition that they are still able to maintain a fully bonded organization upon confinement. The resulting effective potentials are relevant for understanding the behavior of complex mixtures in crowded environments, but may also be exploited for tuning colloidal self-assembly at preferred target distances in order to build desired superstructures.
\end{abstract}

\maketitle

\section{Introduction}
One of the main goals of soft matter physics is to control and tune the self-assembly of colloidal particles into target structures with specific functionalities. There are several routes that can be explored to tailor self-assembly.
One possibility is to directly manipulate colloids by tuning their shape\cite{glotzer2007anisotropy, donev2004unusually,kuijk2012phase, kraft2012surface} or by designing directional and specific interactions\cite{bianchi2006phase,nykypanchuk2008dna,chen2011directed,reinhardt2014numerical, auyeung2014dna,wang2015crystallization, bomboi2016re}. Another popular approach is to indirectly act on colloidal interactions by generating an effective force that is mediated by the suspension in which colloids are immersed. 
A notable example of this situation is the well-known depletion interaction, which arises in solutions of  colloids and non-adsorbing polymers or surfactants (i.e. the cosolute). In this case, an entropy-driven effective attraction is induced among the colloids, deeply altering their behaviour.  
By treating colloid-colloid and colloid-cosolute interactions as hard spheres (HS) ones, while considering polymers ideal, Asakura and Oosawa have shown\cite{asakura1958interaction} that the excluded volume interactions between colloid and cosolute make it favourable for the colloids to be close together. This effective attraction is entirely controlled by the properties of the cosolute, since the range and strength are set by the size and the density of the cosolute respectively. Yet, a slight change of the colloid-cosolute interaction can strongly change this simple, idealized picture\cite{sapir2014origin,rovigatti2015soft}. Nonetheless, the manipulation of colloidal suspensions by adding cosolute is still nowadays one of the most established tools to study the behavior of short-ranged attractive systems\cite{anderson2002insights}. 

It is interesting to note that depletion forces may also play an often underappreciated role in biological systems\cite{marenduzzo2006depletion} as, for example, in the cytoplasm, that is a very dense solution, composed of a large number of macromolecules of different sizes as well as smaller species.  Under these conditions of so-called macromolecular crowding, a confining effect by the large macromolecules affects the medium composed by the smaller species. Such scenario occurs also in several other situations, including clustering of red blood cells in the presence of proteins\cite{faahraeus1929suspension} or the precipitation of tobacco viruses in solution with heparin\cite{cohen1942isolation}. In all these cases it was noticed that the concentration and the type of the cosolute are relevant for the aggregation of large macromolecules\cite{lekkerkerker2011colloids}.

More recently, other types of effective interactions have been investigated, in which the interacting properties of the cosolute play a fundamental role.  The most extreme case is that of a cosolute (or solvent) close to a second-order critical point, which determines the so-called critical Casimir forces. These forces, which can be either attractive or repulsive, were first predicted by Fisher and De Gennes\cite{fisher} and have been recently measured in experiments\cite{hertlein,paladugu,martinez2017energy}. Notably, the range of the interactions is controlled by the thermal correlation length of the cosolute, which diverges at the critical point. A similar mechanism has been shown to produce long-range effective interactions when the cosolute is close to a chemical gelation point\cite{gnan2014casimir}, in the presence of a thermal gradient\cite{kirkpatrick2015nonequilibrium}, in granular materials\cite{cattuto2006fluctuation}, in cell membranes\cite{machta2012critical} and also when the cosolute is composed by active biological systems such as bacteria\cite{ray2014casimir,ni2015tunable}.

Another interesting scenario occurs by considering a cosolute made of molecules which self-associates into chains or larger aggregates, and that can be modeled  with patchy particles with valence two or higher. Two-patch particles form reversible chains and do not experience either phase separation or percolation in the whole phase diagram\cite{rubinstein2010polymer,sciortino2007self}. Such polymerizing cosolute has been adopted in a pioneering experiment\cite{knoben2006long}, which has measured the effective interactions of colloids immersed in a sea of semi-flexible associating supramolecular chains, finding that the effective potential $V_{eff}$ can be attributed to depletion effects with a range that is controlled by the average size of the chains.  To this end, in the experiments the total density of monomers was kept fixed while the concentration of so-called chain stoppers was varied, modulating both the chain length and the range of $V_{eff}$. Very recently, a simulation study of effective potentials of (nano)particles immersed in a solution of (not self-assembling) semi-flexible polymers has been reported\cite{jpc2016}. In this work, the variation of effective interactions induced on the nanoparticles was obtained by tuning the flexibility and the density of the chains while fixing their length. The resulting effective potentials were also associated to depletion effects. Only if attraction between nanoparticles and chains were introduced, the resulting potentials showed more complex features related to bridging phenomena.
Regarding higher valence patchy particles, calculations of effective potentials have been performed by some of us, with the aim to identify the crucial role of the lifetime of the clusters in determining the colloid-colloid interactions\cite{gnan2016}. 

In this article, we perform extensive numerical simulations to calculate the effective interactions between two colloids immersed in a sol of self-assembling patchy particles of different valence. For two-patch particles, we are able to explore the whole phase diagram at a variety of different conditions, exploiting the absence of phase separation. Beyond depletion effects, we identify a novel regime for $V_{eff}$, which develops an oscillatory behavior, still being totally attractive. This occurs thanks to the ability of the chains to be efficiently confined by colloids, optimizing at the same time both the energy and the entropy contributions in the free energy, through the onset of local nematic order. We thus generalize this argument and identify conditions by which higher valence particles, in particular three- and four-patch particles, also show the same behavior. The resulting effective potentials, displaying local attractive minima at particular distances that can be tuned varying the size ratio between colloids and cosolute, are thus highly appealing to drive colloidal self-assembly into preferred structures. 

\section{Model and methods}
\begin{figure}[t]
\centering \includegraphics[width=8.7 cm]{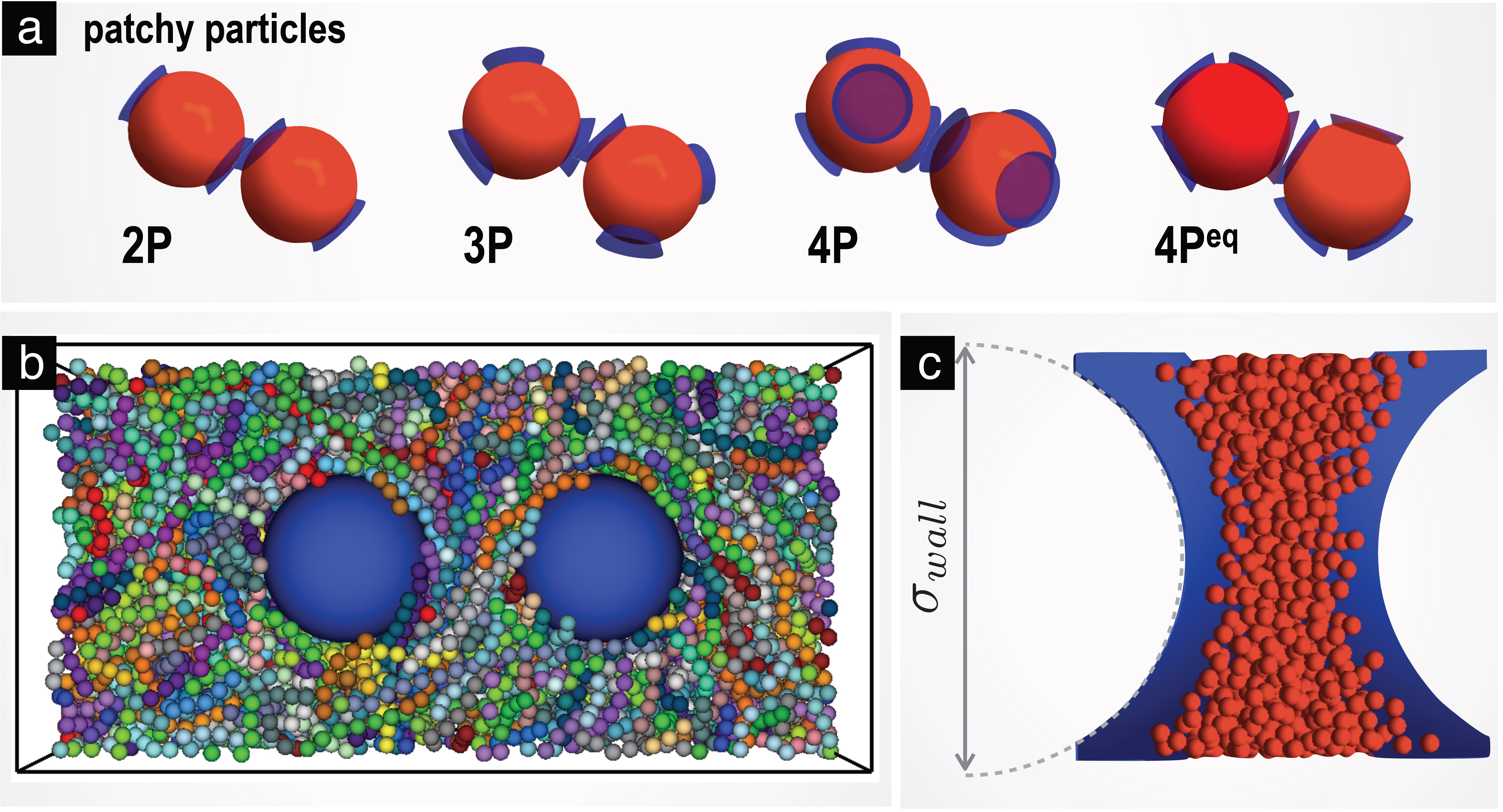}
\caption{(a) Illustration of the types of patchy particles simulated  and their patch arrangements;
(b) Snapshot of a Monte Carlo NVT simulation of two colloids (blue spheres) in solution with 2P patchy particles at $\phi=0.262$ and $T=0.100$. Different colours indicate chains of different lengths;
(c) Snapshot of the Grand Canonical Monte Carlo 
%$\mu$VT 
simulation of patchy particles (red) confined between two spherical walls (blue) of size $\sigma_{wall}$.} 
\label{fig1new}
\end{figure}

To achieve the assembly of monomers into reversible chains and clusters, we model cosolute particles as hard-spheres (HS) of diameter $\sigma_m$ decorated with attractive sites (patches) interacting with the Kern-Frenkel potential\cite{Kern2003fluid}. The potential consists of a square-well attraction of width $\delta$ and depth $\varepsilon$ modulated by an angular function that depends on the mutual orientation of the particles and on $\cos(\theta_{max})$ that accounts for the volume available for bonding.  We fix $\delta=0.119\sigma_m$ and $\cos(\theta_{max})=0.894717$ in order to guarantee the one-bond-per-patch condition\cite{giacometti2010effects}. We consider patchy particles of different valence and patch arrangements, namely 2P particles with two patches located on the poles, 3P particles with three patches symmetrically located on the equator, particles with four patches arranged either in a tetrahedral order (4P) as well as aligned on the equatorial plane (4P$^{eq}$)\cite{bianchi2006phase,bianchi2008theoretical}. The different models are illustrated in Fig.~\ref{fig1new}(a).

In all cases, we evaluate $V_{eff}(r)$ between two HS colloids of size $\sigma_c$ immersed in a solution of patchy particles by performing Monte Carlo (MC) simulations in the canonical ensemble. The system contains a number of particles ranging from $N=10279$ to $N=10870$ monomers and the size ratio between these and HS colloids is fixed to $q=\sigma_m/\sigma_c=0.1$. We evaluate $V_{eff}(r)$ along the $x$-direction ($r=x$), hence colloids are constrained to move only along the $x$-axis while cosolute particles can explore the simulation box in all directions. In the following, $r$ is used to indicate the surface-to-surface distance between the colloids. The simulation box is chosen to be parallelepipedal with $L_x$ ranging from $42\sigma_m$ to $76\sigma_m$, depending on the density of the system, and $L_y=L_z=L_x/2$. With this choice we ensure that $r$ is always larger than the distance at which $V_{eff}$ goes to zero.  Periodic boundary conditions are applied in all directions.  We evaluate the probability $P(r)$ to find the two colloids at a given distance $r$ through an umbrella sampling technique\cite{gnan2012tuning}, so that $\beta V_{eff}=-ln(P(r))$, where $\beta=1/(k_B T)$ with $k_B$ the Boltzmann constant and $T$ the temperature, besides a constant that is set by imposing $V_{eff}(\infty)=0$. The cosolute diameter $\sigma_m$ and the well depth $\epsilon$ are chosen as units of length and energy, while temperature is measured in units of $\epsilon$ (i.e. $k_B=1$).

In the case of 2P cosolute particles, we calculate $V_{eff}$ in a wide region of temperatures and packing fractions $\phi=\pi\rho\sigma_m^3/6$, with $\rho$ being the number density, taking advantage of the absence of phase separation. A low $T$ simulation snapshot for 2P cosolute is shown in Fig.~\ref{fig1new}(b). For 3P, 4P and 4P$^{eq}$ particles, we focus on isochores $\phi=0.262$ and $\phi=0.314$. The calculations are performed at the critical temperatures of the models, i.e. $T_c^{3P}=0.125$ and $T_c^{4P}=0.168$ \cite{foffi2007possibility} for 3P and 4P cosolute, while for 4P$^{eq}$ we have roughly estimated from Grand Canonical simulations  that the critical temperature is $T_c^{4P^{eq}}\approx 0.148$, and simulations have been performed at $T=0.147$.

We also perform Grand Canonical Monte Carlo (GCMC) simulations of cosolute particles of various types (2P, 3P, 4P and 4P$^{eq}$) between two spherical walls, which mimic the surfaces of the two colloids. In this way we are able to assess the effect of confinement operated by the colloids and to visualize the geometrical arrangement of the cosolute between the two colloids. To this aim, we use a simulation box of $[20+\Delta_x \sigma_m, 20\sigma_m,20\sigma_m]$ where $\Delta_x$ is varied in order to span different distances between the walls. Periodic boundary conditions are applied only along $y$ and $z$-axis. The two walls are obtained by centering two spheres of size $\sigma_{wall} > L_y$  in the simulation box along $x$  and are placed at a surface-to-surface distance $\Delta_x$ along the $x$-axis. The curvature of the walls is varied by simply changing the size of the two spheres. An illustration of the GCMC simulations in the confined geometry is provided in Fig.~\ref{fig1new}(c).  In order to estimate the curvature effect, we consider both $\sigma_{wall}=30\sigma_m$ and $\sigma_{wall}=500\sigma_m$. GCMC simulations are performed at the same temperature as the NVT simulations and at a value of the chemical potential $\mu$ which depends on the type of cosolute employed. To estimate its value we have initially performed bulk GCMC simulations of cosolute particles (in the absence of the walls) at constant $T$ for various $\mu$ until we have found the same average packing fraction as that of the NVT simulations. Thus, GCMC simulations in the presence of spherical walls are carried out for 2P particles at $(\mu=0.050, T=0.100)$, for 3P particles at $(\mu=0.030, T=0.125)$, for 4P particles at $(\mu=0.035, T=0.168)$ and for 4P$^{eq}$ at $(\mu=0.015,T=0.147)$.

\section{Results}
\subsection{Effective potentials generated by 2P particles}
We start by reporting results for the effective potential arising between two colloids immersed in a solution of 2P particles, which form reversible polymer chains
whose length is controlled by $T$ and $\phi$ of the monomers\cite{sciortino2007self}.  The chains flexibility is fixed by patch-patch interactions. 
For our choice of model parameters, the persistence length of the chains, calculated as in\cite{LpBinder}, is $\simeq 10.3\sigma_m$. This value is considerably smaller than that of the experiment of Ref.\cite{knoben2006long} but still relatively large, approaching the semi-flexible regime.

\begin{figure}[t]
\centering \includegraphics[width=8.7 cm]{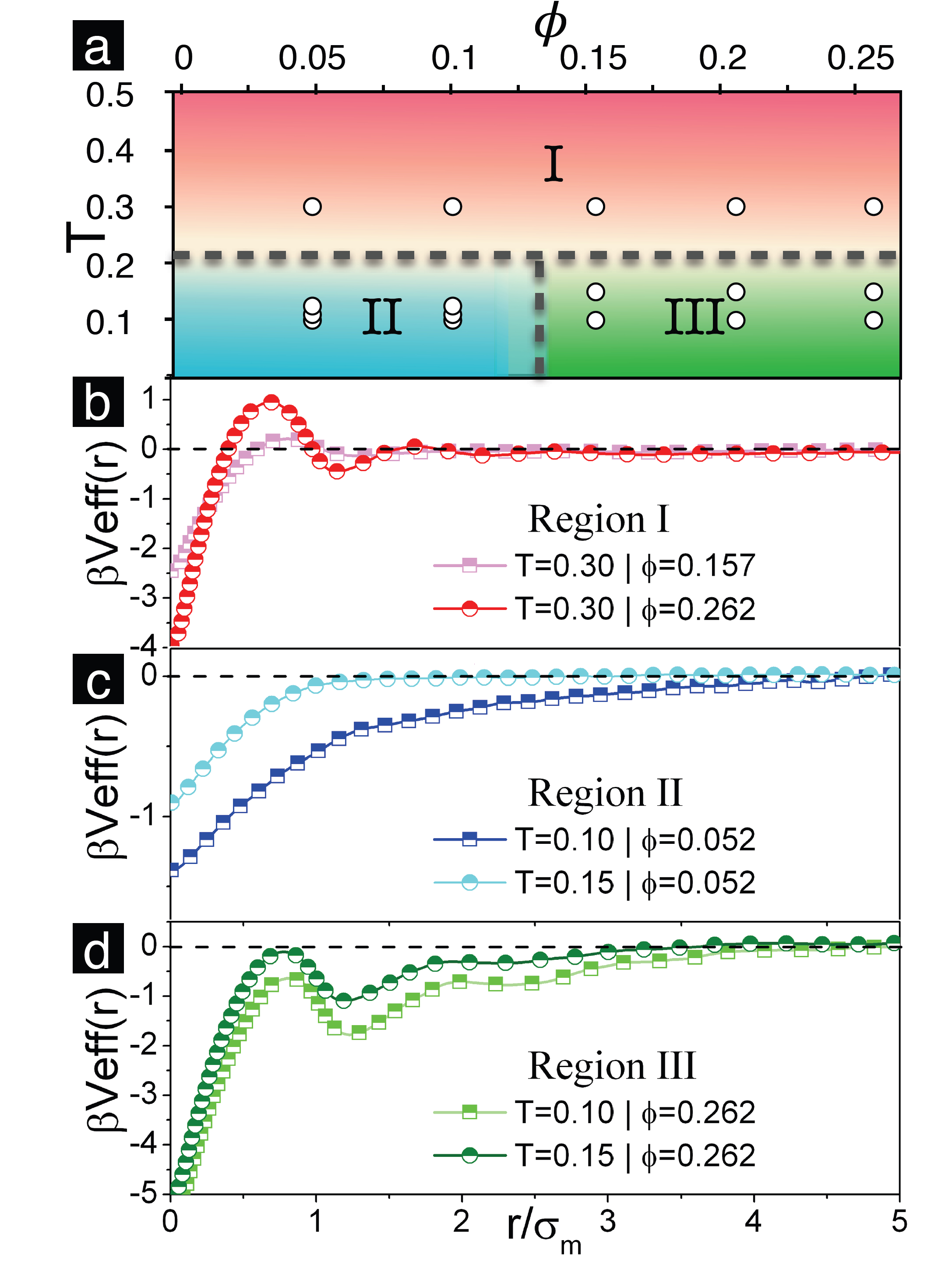}
\caption{(a) Investigated 2P phase diagram highlighting the three regions associated to different behaviours of $V_{eff}$; (b)-(d) Effective potentials $\beta V_{eff}$ calculated for selected state points in the three regions shown in (a).
}
\label{fig2}
\end{figure}

We perform calculations for a wide variety of state points, shown in Fig.~\ref{fig2}(a), exploring packing fractions $\phi$ in the range $(0, 0.3)$ and temperatures $T$ in the interval $(0.1, \infty)$. We find that the resulting $V_{eff}(r)$ can be classified in three distinct regions of the phase diagram, illustrated in Fig.~\ref{fig2}(a), according to three different behaviours represented in Fig.~\ref{fig2}(b-d).

In region (I), which amounts to high temperatures ($T\gtrsim 0.3$) for all investigated packing fractions, thermal fluctuations are dominant with respect to the ability of monomers to form bonds. Thus the 2P monomers behave mostly as HS particles and the colloids experience the typical potential observed in highly asymmetric hard-sphere mixtures\cite{dijkstra1999phase}: an initial attraction at short distances, followed by oscillations, both repulsive and attractive, that fade away at large colloid distances. These oscillations are associated to a layering of particles between the colloids and have been previously observed experimentally and numerically in several works \cite{crocker1999entropic,bechinger1999understanding,dijkstra1999phase,gotzelmann1999depletion}.

In region (II), corresponding to low temperatures ($T\lesssim 0.3$) and low packing fractions ($\phi\lesssim 0.15$), the cosolute particles are able to exploit the self-assembly into chains and thus $V_{eff}$ turns completely attractive and its range increases with decreasing $T$. This can be associated to a progressive increase of the average chain length at low temperatures, in good agreement with experiments in Ref.\cite{knoben2006long}. In this regime, a depletion mechanism is thus responsible for the observed behavior of $V_{eff}$ where chains, rather than monomers, act as depletants. 

In region (III) the resulting $V_{eff}$ shows novel, interesting features: it is still completely attractive, but also develops oscillations similar to those observed in the HS-like regime. 
Since this behavior contains both features observed in the other two regions, it is tempting to think that it results from the sum of the two effects: confinement due to excluded volume and depletion of chains. The non-monotonicity of the potential further suggests that there exist special distances where attraction is enhanced or depressed, that might have to do with the organization of monomers and/or chains in between the colloids, similarly to the case of HS-induced depletion. However, a crucial question remains: why is the potential totally attractive?

\begin{figure}[t]
\centering \includegraphics[width=8.7cm]{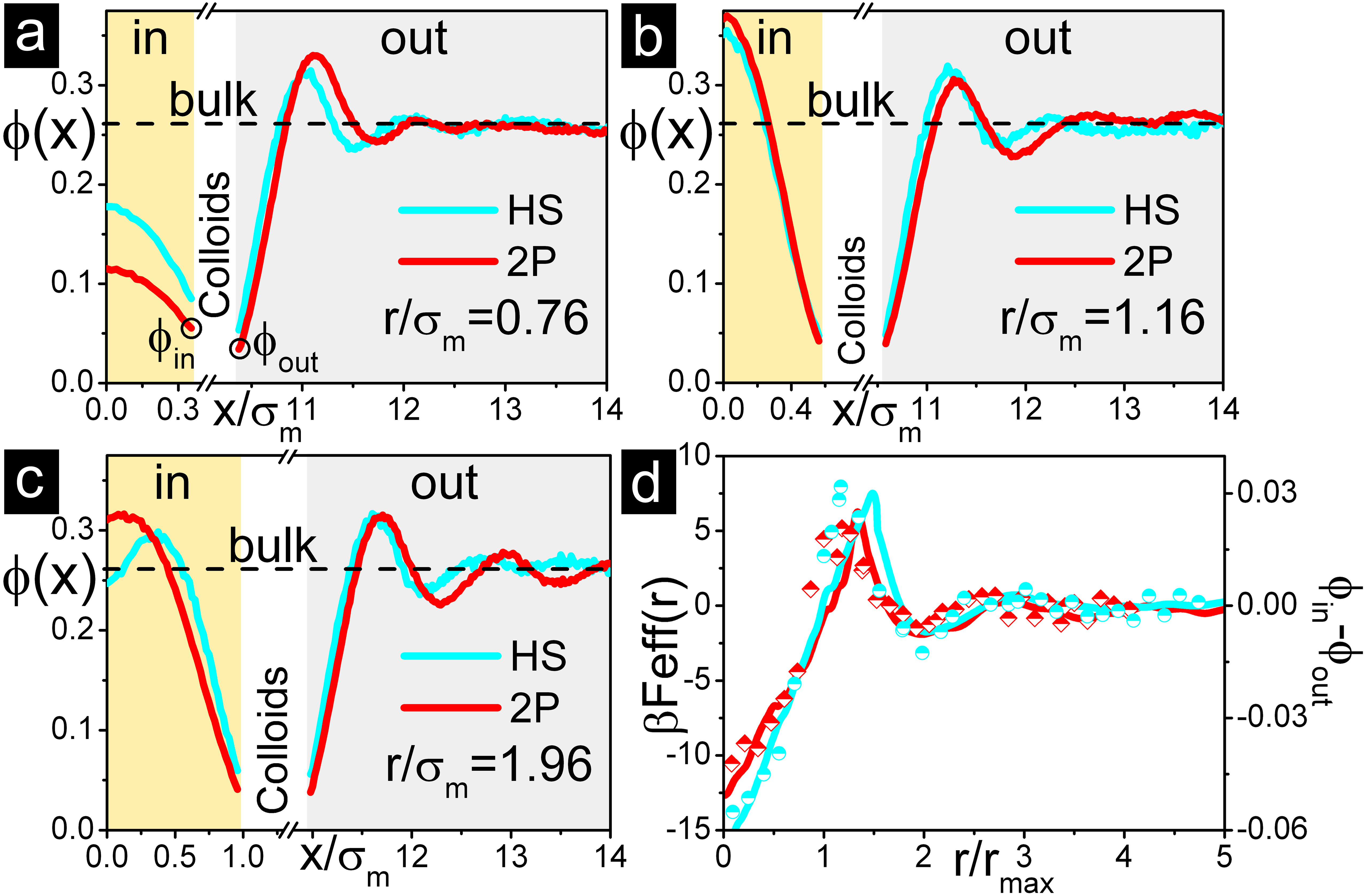}
\caption{Packing fraction profiles $\phi(x)$ of the monomers along the $x$-direction at three different colloid distances $r/\sigma_m$, corresponding roughly to (a) the first local maximum, (b) the first local minimum and 
(c) the second local maximum in $V_{eff}$ for both HS and 2P monomers. For the latter $T=0.100$. Due to symmetry only half of the profile is shown:
labels `` in'', ``out''  indicate the region between and on one side of a colloid respectively. Profiles are calculated in a sub-volume of the simulation box of size $[L_x, 3\sigma_m, 3\sigma_m]$; 
(d) the effective force $\beta F_{eff}$ (lines) and the packing fraction mismatch $\phi_{in}-\phi_{out}$ at contact (symbols) as a function of $r/r_{max}$. 
}
\label{fig3}
\end{figure}

To answer this question, we need to identify the microscopic mechanism leading to the behavior of $V_{eff}$. We start by examining the role of depletion, here intended as exclusion of the cosolute from the region between the colloids. We then calculate the packing fraction profiles $\phi(x)$ of the monomers along the $x$-direction for different colloidal distances $r$, comparing results for the 2P case at $T=0.1$ and for the HS system at the same cosolute packing fraction ($\phi=0.262$). This is shown in Fig.~\ref{fig3}(a-c).  We find that in the region between the colloids the packing fraction of the monomers is lower than in the bulk for colloid distances smaller the first local maximum of the potential, that we call $r_{max}$. For $r > r_{max}$, monomers population increases in between the colloids, exceeding the bulk value of $\phi$. This trend persists also at larger distances, finding no distinctive signatures in correspondence to the subsequent maxima and minima of $V_{eff}$. Most strikingly, these results are qualitatively similar for HS and 2P cosolutes, despite a systematic shift of the oscillations of the 2P data towards larger distances. Indeed, if we compare $\phi(x)$ at the same $r$, we find that the formation of layers is always anticipated by HS with respect to 2P monomers. 
Thus, for example while HS start to form a second layer in the region between the colloids, 2P are still forming the first one, as shown in Fig.~\ref{fig3}(c). This is compatible with the fact that 2P monomers can be considered to have a larger effective size which is controlled by the thermal correlation length as shown already for cluster-forming cosolutes\cite{gnan2016discontinous}. Thus, the layering effect is less efficient at the same distance with respect to HS. 

The similarity of the $\phi(x)$ profiles for HS and 2P cosolute is consistent with the behavior of the effective forces $F_{eff}$, shown in Fig.~\ref{fig3}(d) as a function of $r/r_{max}$  in order to remove the systematic shift between HS and 2P cosolutes.  The force is generated by an osmotic pressure unbalance, that is proportional to the mismatch $\phi_{in}-\phi_{out}$ of the packing fraction at contact, respectively in the regions inside ($\phi_{in}$) and outside ($\phi_{out}$) the colloids. In Fig.~\ref{fig3}(d) the calculated $\phi_{in}-\phi_{out}$ is also reported, again showing no qualitative differences for HS and 2P cosolutes. Interestingly, the force becomes repulsive for distances just above the potential minimum, but the potential is still attractive.
From these results, it is clear that the force, the osmotic pressure unbalance and all the standard depletion arguments are not useful in order to rationalize the novel behavior of $V_{eff}$ found in the presence of 2P monomers with respect to HS ones.

\begin{figure*}[t]
\centering \includegraphics[width=17 cm]{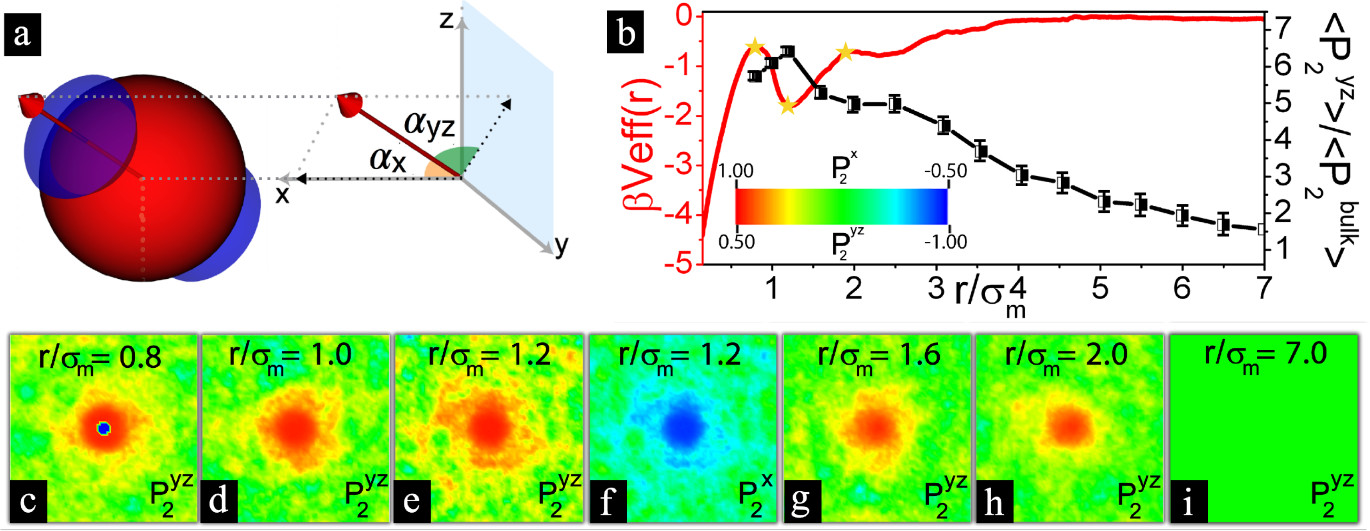}
\caption{(a) Illustration of the angles between the orientation vector of a 2P monomer and the $yz$-plane (or $x$-axis);
(b) $\beta V_{eff}(r)$ for the 2P particles in comparison to the normalized (with respect to the bulk value) $\langle P_2^{yz}\rangle$ averaged over a circular slice, centered at the mid-point between the colloids, of radius $5\sigma_m$ and height $0.1\sigma_m$. (c-e,g-i) maps of $P_2^{yz}$ and (f) map of $P_2^{x}$ calculated in a central rectangular slice of dimensions $[L_x, 0.1\sigma_m, 0.1\sigma_m]$ for representative values of $r/\sigma_m$. The maps are coloured according to the bar provided in panel (b).}
\label{fig4}
\end{figure*}

To gain microscopic understanding, we turn to address how the chains can arrange in between the colloids, maintaining their connectivity. To this aim, it is intuitive to think that they should adopt a preferential alignment along a direction orthogonal to the colloid center-to-center distance. To monitor this effect, we evaluate the nematic order parameter of the monomers with respect to the $yz$-plane, which is the standard Legendre polynomial with a proper normalization to account for the plane projection, which guarantees a zero mean of the parameter, i.e. $P_2^{yz} =\langle \frac{3}{2}cos^2(\alpha_{yz})-1\rangle$, 
where $\alpha_{yz}$ is the angle between the orientation vector of a monomer and the $yz$-plane (see Fig.~\ref{fig4}(a)). The corresponding $P_2^{yz}$ maps are shown in Fig.~\ref{fig4}(c-e,g-i) for several colloid-colloid distances. We find confirmation that particles, and thus chains, become more and more correlated in orientations, showing the onset of nematic order. By averaging these findings over an appropriate sub-volume, we find that the behavior of $\langle P_2^{yz}\rangle $ is non-monotonic and strictly follows the oscillations of the potential, as shown in Fig.~\ref{fig4}(b).  We remark that in each configuration most of the chains are aligned along a preferred direction which is randomly chosen in the $yz$-plane. By monitoring also $P_2^{x}$ with respect to the $x$-axis, defined as $P_2^{x} =\langle \frac{3}{2}cos^2(\alpha_{x})-\frac{1}{2}\rangle$ and shown in Fig.~\ref{fig4}(f), we find that this is instead much lower than the bulk case, confirming that the alignment in the $r$-direction is strongly suppressed. The tendency of the chains to nematize is an echo of the behavior of confined polymers, which at large densities always prefer to order close to a surface \cite{zhang2016surface}. In the present case, the chains are able to satisfy both energy (bonding) and entropy (excluded volume) constraints, thus making the effective potential completely attractive.  

It is now important to compare our results with the recent calculations of $V_{eff}$ between two (nano)-particles in the presence of (non-self-assembling) polymer chains of a given length and flexibility, reported in \cite{jpc2016}. When the two particles interact with the chains via excluded-volume only as in our model,  the effective potentials are found to be monotonically attractive for all investigated densities and values of the flexibility, similarly to what we observe in Fig.~\ref{fig2}(b) for the depletion-like regime. 
These findings confirm that an attractive, non-monotonic behavior of $V_{eff}$ cannot be obtained with a cosolute which does not self-assemble, because the crucial competition between entropy and energy is not active in this case, so that depletion is always dominant\cite{note}. We thus conclude that the  self-assembly of the cosolute is crucial in this setup to give rise to attractive oscillatory potentials.

\subsection{Oscillatory attractive effective potentials generated by 3P and 4P particles}
The behaviour of $V_{eff}$ induced by 2P particles in regions I and II has been already observed for several cosolute models, independently if they are anisotropic (e.g. 3P)\cite{gnan2012tuning,gnan2016discontinous} or isotropic (e.g. square-well (SW))\cite{GnanSM2012} and is attributable to depletion effects, in agreement with experimental studies for both HS depletants\cite{crocker1999entropic,bechinger1999understanding} and added chains\cite{knoben2006long}. We thus focus in the following only in the results found in region III. In order to provide generality to the 2P cosolute results and to determine the minimal ingredients needed to observe effective oscillatory attractions, we also investigate different types of self-assembling patchy particles acting as cosolute.

\begin{figure}[t]
\centering \includegraphics[width=8.7 cm]{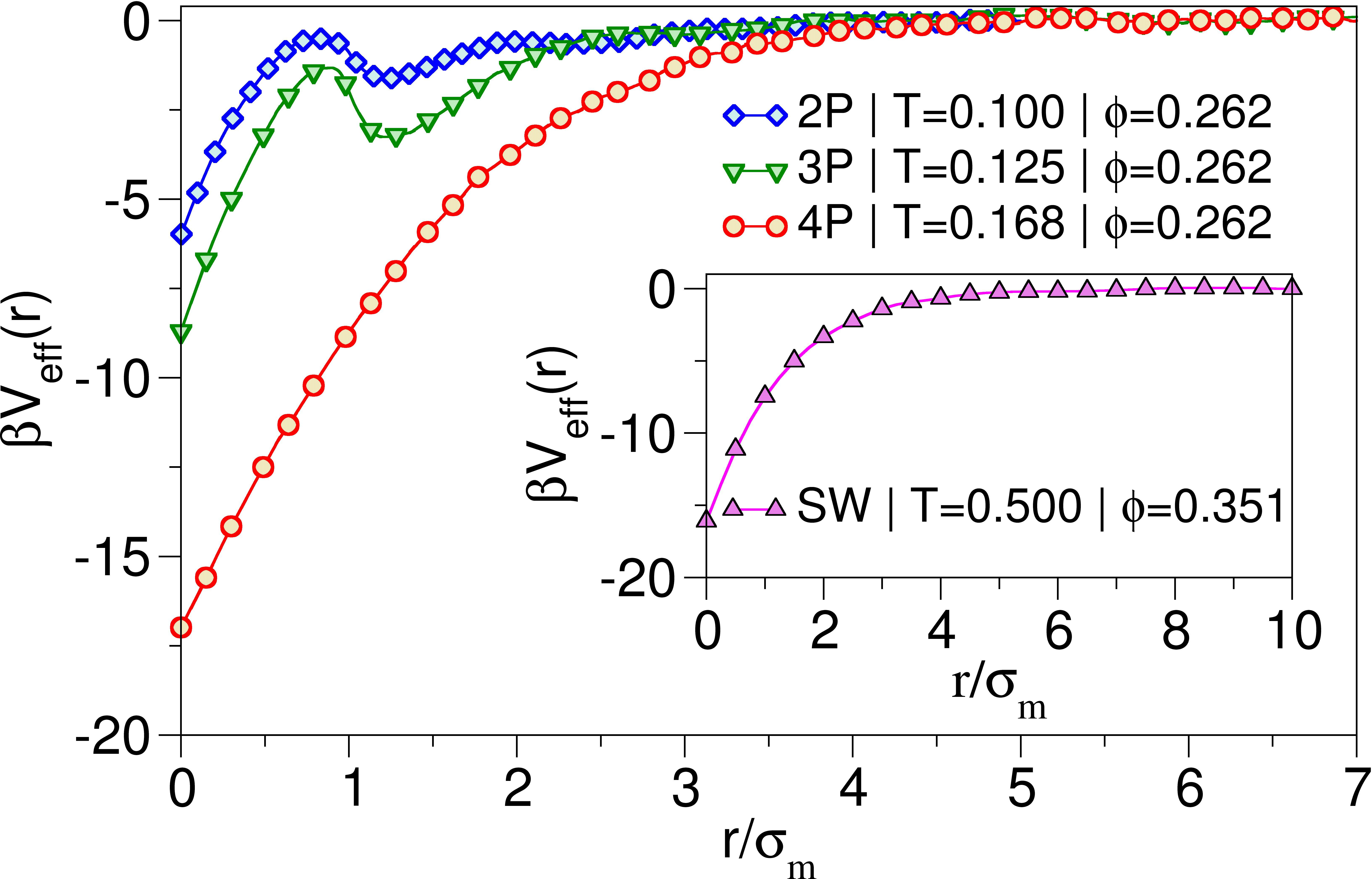}
\caption{Effective potential $\beta V_{eff}$ for different types of cosolute particles with $q=0.1$ at $\phi=0.262$. For 3P and 4P particles, $V_{eff}$ has been evaluated at the critical temperature $T_c^{3P}=0.125$ and $T_c^{4P}=0.168$, respectively. For the 2P cosolute, we refer to the lowest investigated temperature ($T=0.100$); Inset: $\beta V_{eff}$ obtained in the case of a SW cosolute of $10\%$ width for $\phi=0.351$ and $T=0.500$, the latter being slightly larger than the critical temperature\protect\cite{GnanSM2012}. }
\label{fig5}
\end{figure}

In the case of 2P cosolute we have found that an unusual, oscillatory attraction arises because the cosolute is able to preserve the maximally bonded pattern upon confinement. Comparing this behavior with that obtained in the case of generically attractive additives (e.g. isotropic square-well cosolute) which form branched, extended clusters, we see that for the latter only a standard, monotonic depletion-like attraction is found, even at high densities\cite{GnanSM2012}, because the confinement disfavours the formation of clusters. Thus no preferential distances exist and no minima are found in the effective potential.
It is therefore legitimate to ask  whether anisotropic interactions play a role in determining the oscillatory attractive behavior of the effective interaction at high density. To this aim we compare the effective potential generated by 2P and SW particles in the high density regime with that of 3P and 4P particles, in order to fill the gap between the behaviour of a highly anisotropic cosolute and a spherically-interacting one.  The effective potentials arising between two colloids in a dense solution ($\phi=0.262$) of different cosolutes are shown in Fig.~\ref{fig5}. We find that $V_{eff}$ mediated by the 3P particles still displays an attractive oscillatory character, while for the 4P ones no oscillations are found, as also seen in the case of isotropically attractive cosolute \cite{GnanSM2012}.  

\begin{figure*}[t]
\centering \includegraphics[width=\textwidth]{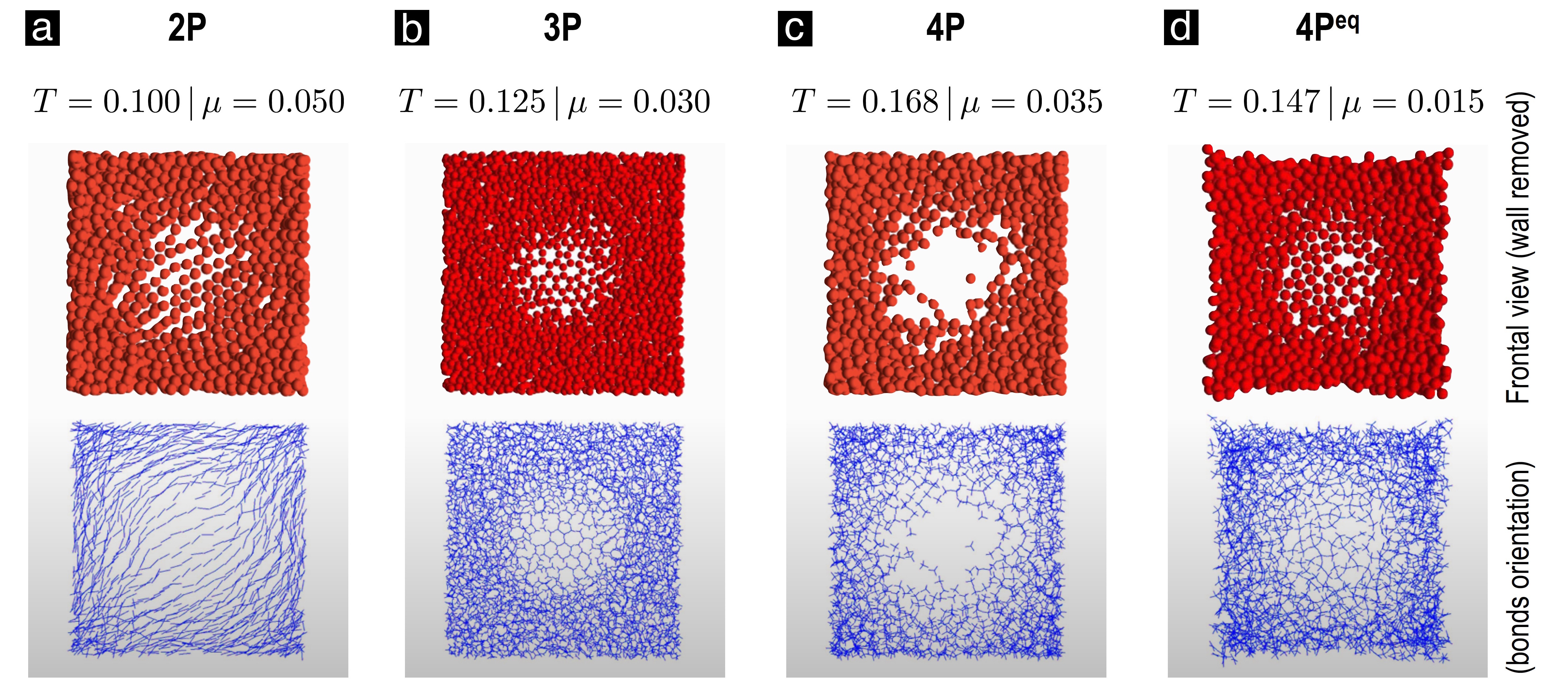}
\caption{ (a) 2P, (b) 3P , (c) 4P and (d) 4P$^{eq}$ simulations between two spherical walls with $\sigma_{wall}=30\sigma_m$ at distance $\Delta_x=1.19\sigma_m$: top panels show the frontal view of the simulation box (the two walls have been removed to allow the observation of  the confinement-assisted assembly in the middle of the simulation box); bottom panels illustrate the bond view of the particle configurations in the middle panels. For $2P$,$3P$ and $4P$ particles, the average packing fraction is $\phi=0.262$ and temperatures are the same as in Fig.~\protect\ref{fig5}. For $4P^{eq}$ particles, the average packing fraction is $\phi=0.314$ and $T=0.147$. %Particles with patches arranged on the equatorial plane are still able to form bonds upon confinement.
}
\label{fig6}
\end{figure*}

\begin{figure}[t]
\centering \includegraphics[width=8.7 cm]{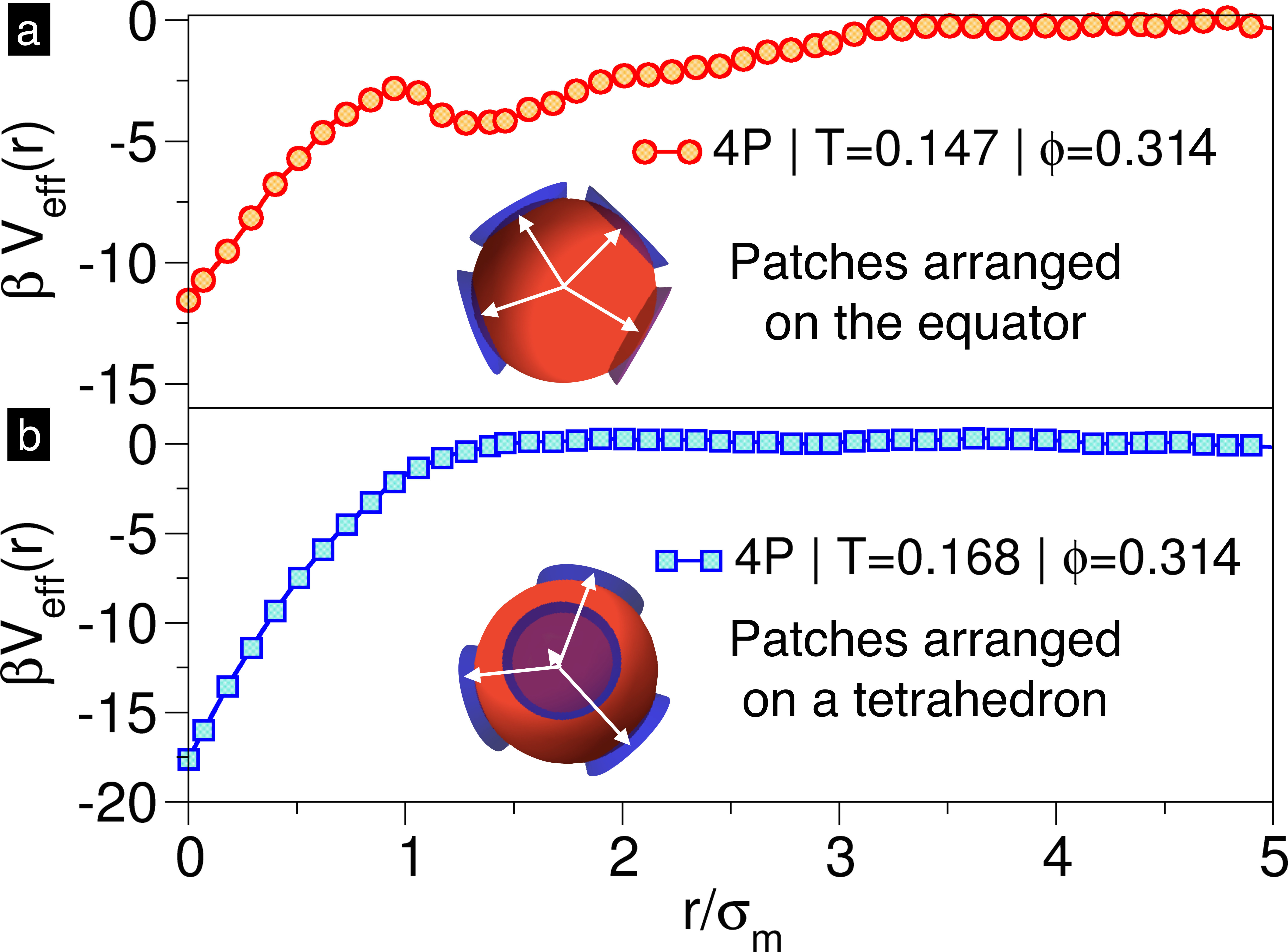}
\caption{Effective potential calculated for $4P$ cosolute in the case of  (a) patches arranged on the equatorial plane and  (b) on a tetrahedron at $\phi=0.314$. The temperatures are close to the bulk critical temperatures of the two models.}
\label{fig7}
\end{figure}

In order to better grasp the microscopic mechanism of confinement between the surfaces of the two colloids, we rely on Grand Canonical MC simulations of patchy particles between two spherical walls that mimic the surface of the two colloids.  In Fig.~\ref{fig6} we report snapshots of the cosolute particles between  spherical walls at a distance corresponding to that of the first minimum of the effective potential. A striking difference in the behavior of 2P and 3P particles with respect that of 4P particles is observed. While 2P and 3P particles are found also in the region between the colloids (a,b middle panels), 4P particles are not (c, middle and lower panel). This is due to the fact that both 2P and 3P are able to occupy the plane in between the colloids in a fully bonded pattern (Fig.~\ref{fig6} (a) and (b) lower panels), because 2P particles form chains, which are tangential to the colloids surface, while 3P ones can form a planar structure which still can  fit into the confined region. On the other hand, 4P particles prefers to form a tetrahedral network, whose 3D structure cannot efficiently occupy the reduced space between the colloids and thus, they prefer to remain outside of this region, giving rise to simple depletion. This is confirmed by the larger value at contact observed for 4P tetrahedral cosolute with respect to lower valence particles at the same density. These results are enhanced in the case of almost planar walls $\sigma_{wall}=500\sigma_m$ (not shown). 
We thus find that these results fully support our interpretation of the microscopic mechanism underlying the occurrence of an effective oscillatory attractive potential, which arises when the cosolute is still able to efficiently bond upon confinement.  A final test to this interpretation can be made by employing 4P$^{eq}$ particles whose patches are arranged on the equatorial plane, which then prefer to form planar rather than tetrahedrally-bonded structures. We thus repeat the calculation of $V_{eff}$ for this modified patch geometry, whose results are reported in Fig.~\ref{fig7}(a). In the case of 4P$^{eq}$ particles, we do indeed recover the oscillatory attractive behavior, which is never found in the case of tetrahedrally organized patches. Performing also GCMC simulations of 4P$^{eq}$ particles in between spherical walls, we find confirmation that  particles with patches aligned onto the equatorial plane recover the ability to efficiently form fully-bonded patterns in between the colloids surface, as shown in Fig.~\ref{fig6}(d).

In summary, the employment of limited-valence cosolute which is able to efficiently order in the confined space between the colloids gives rise to a novel kind of attractive effective potentials, which shows the presence of maxima and minima at preferred distances. The location of these maxima and minima is set by the colloid-cosolute size ratio. The ability to self-assemble of the cosolute is a necessary ingredient for the observed behavior, which does not occur in the case of fixed polymer chains\cite{jpc2016}. Indeed, there exist special distances at which self-assembling patchy particles are able to fit in the confining region in a fully-bonded way e.g. by locally order either into chains in a quasi-nematic fashion for 2P particles or into planar structures for 3P and 4P$^{eq}$ ones. Such distances, compatible with the size of the individual cosolute particles as shown in Fig.~\ref{fig5}, thus identify the location of maxima and minima in the resulting $V_{eff}$.

\section{Conclusions}
In this work we have proposed a strategy to tune and control colloidal self-assembly by adding complex additives in the suspension. In particular, we have shown that the use of patchy particles as cosolute provides a viable route to produce non-monotonic attractive effective potentials between the colloids.

We first evaluated $V_{eff}$ between colloids immersed in a cosolute made of non-adsorbing, self-assembling polymer chains.  In addition to standard depletion at low densities and HS behavior at high temperatures, we have identified a new behavior of $V_{eff}$ occurring at high density and high attraction strengths of the monomers. The resulting effective interaction consists of an intriguing oscillatory attraction that we have explained as the result of the ability of the assembled chains to be confined by the colloids which allows them to optimize simultaneously bonding and excluded volume constraints.  It is important to stress that this behavior is not observed when the cosolute is made of (repulsive) polymer chains which do not self-assemble, as clearly shown in Ref.~\cite{jpc2016} (see in particular Figs. 1 and 2), because in this case no competition between bonding and entropy takes place and depletion always wins.

We have then generalized our findings for cosolutes made of higher valence particles, finding that for 3P patchy particles this tendency is confirmed and even enhanced, because the particles are still able to form fully-bonded planar structures in the confined region between the colloids. However, using 4P particles that form a tetrahedral network completely suppresses the presence of oscillations in the potential, which thus reduces to the standard, depletion-driven, attractive potential also observed for isotropically attractive cosolutes. Thus, a necessary condition to observe  non-monotonic attractive potentials is that the cosolute particles are able to be confined by the colloids maintaining the ability to fully bond. This is a very simple conceptual situation which may have important consequences in the context of materials design.  Indeed, the presence of multiple lengthscales due to the existence of attractive, local minima and maxima in $V_{eff}$ paves the way for a rational design of mixtures of colloids and different types of cosolutes which could result in the assembly of target colloidal structures, such as low-coordinated crystals \cite{rechtsman2006self,edlund2011designing,marcotte2013communication,jain2014dimensionality}, quasicrystals\cite{ryltsev2015self,engel2015computational,metere2016smectic,glotzer2017non} and other complex scenarios\cite{jadrich2015equilibrium,lindquist2017interactions}.
While non-monotonic potentials have been extensively studied, especially in the context of inverse design\cite{torquato2009inverse,cohn2009algorithmic,jain2014inverse}, no experimental route has been proposed so far. Our work shows how it is possible to obtain such potentials experimentally, even for colloids whose direct interactions are strictly isotropic.  Thus, colloidal suspensions immersed in a solvent plus a patchy cosolute could represent a suitable model system to realize these potentials in the laboratory. The optimization of the several involved parameters, particularly the colloid/cosolute size ratio, but also the mixture composition, charge, etc. will be crucial to identify and obtain a desired target structure. The possibility to extend this study to the case of a binary cosolute appears to be particularly promising to generate different oscillations at competing distances and to vary ad-hoc the amplitude of these oscillations as well as their nature (attractive or repulsive). Such optimization, beyond the scope of the present manuscript, will be tackled in future work.

Finally, we expect that our results will be quite relevant in situations of biological interest\cite{marenduzzo2006depletion,sapir2014origin,sukenik2013balance}. Indeed, they apply in crowded environments, where cosolute particles are confined by colloids in very dense, complex solutions. Recent works have also reported that food proteins experiencing depletion interactions behave in a similar way as colloidal particles\cite{bhat2006spinodal,van2012origin,mahmoudi2015making}, suggesting that our study could be also exported to induce non-monotonic attractive potentials for systems such as proteins and other (bio)-macromolecules. 

\section*{Conflicts of interest}There are no conflicts of interest to declare.

\section*{Acknowledgments}
We acknowledge support from ERC Consolidator Grant 681597 MIMIC (NG, EZ) and from MIUR `Futuro in Ricerca' ANISOFT/RBFR125H0M (NAG, EZ).

\bibliography{NematicBib}

\end{document}